\documentclass[fleqn,12pt,twoside]{article}
\usepackage{espcrc1}

\usepackage{graphicx}
\usepackage[figuresright]{rotating}

\usepackage{epsfig}
\newcommand{\AmS}{{\protect\the\textfont2
  A\kern-.1667em\lower.5ex\hbox{M}\kern-.125emS}}

\title{An approach toward the successful supernova explosion 
by physics of unstable nuclei}

\author{K. Sumiyoshi\address{Numazu College of Technology,\\ 
        Ooka 3600, Numazu, Shizuoka 410-8501, Japan}\thanks{E-mail: sumi@numazu-ct.ac.jp, K.S. 
        expresses special thanks for the usage of VPP supercomputers of RIKEN, NAO and JAERI in Japan.},
        S. Yamada\address{Science and Engineering, Waseda University,\\
        Ohkubo 3-4-1, Shinjuku, Tokyo, 169-8555, Japan},
        H. Suzuki\address{Faculty of Science and Technology, Tokyo University of Science,\\
        Yamazaki 2641, Noda, Chiba 278-8510, Japan},
        H. Shen\address{Department of Physics, Nankai University,\\
        Tianjin 300071, China}
        and
        H. Toki\address{Research Center for Nuclear Physics (RCNP), Osaka University,\\
        Mihogaoka 10-1, Ibaraki, Osaka 567-0047, Japan}
        }
\begin{document}

% typeset front matter
\maketitle

\begin{abstract}
We study the explosion mechanism of collapse-driven supernovae 
by numerical simulations with a new nuclear EOS based on unstable 
nuclei.  
We report new results of simulations of general relativistic 
hydrodynamics together with the Boltzmann neutrino-transport 
in spherical symmetry.  
We adopt the new data set of relativistic EOS 
and the conventional set of EOS (Lattimer-Swesty EOS) 
to examine the influence on dynamics of core-collapse, bounce 
and shock propagation.
We follow the behavior of stalled shock more than 500 ms 
after the bounce and compare the evolutions of supernova core.

\end{abstract}

\section{Introduction}

Understanding the explosion mechanism of core-collapse supernovae 
is a challenging problem, that requires extensive researches 
in nuclear physics and astrophysics.  
In order to reach the final answer, it is necessary to investigate 
the core-collapse supernovae by implementing hydrodynamics and 
neutrino-transfer together with reliable nuclear equation of states 
and neutrino-related reactions.  
In this regard, recent numerical simulations of neutrino-transfer 
hydrodynamics \cite{Ram00,Mez01,Lie01,Tho03}
have cast a light to the importance of nuclear physics 
as well as neutrino-transfer, though they show no explosion at the moment.  
Meanwhile, advances in physics of unstable nuclei have 
given chances to provide us with nuclear physics in supernovae 
than ever before, therefore, those new nuclear data 
should be examined in modern supernova simulations.  
In this paper, we focus on the influence of the new nuclear equation 
of state (EOS) in the neutrino-transfer hydrodynamics. 
We follow the core-collapse, bounce and shock propagation 
by adopting the new equation of state, which is based on the data of 
unstable nuclei, and the conventional one which has been used 
almost uniquely in recent simulations.  
We compare the behavior of shock and the thermal evolution of 
supernova core by performing numerical simulations for a long 
period of $\sim$1 sec after the core bounce.  
%More recent studies have reported the results of supernova 
%simulations with new electron capture rates (and EOS) 
%to assess the impact on the core-collapse and bounce.  

\section{A new nuclear EOS table}

Recently, a new complete set of EOS 
for supernova simulations (Shen's EOS) has become available \cite{She98a,She98b}.  
The relativistic mean field (RMF) theory with a local density approximation 
has been applied to the derivation of the supernova EOS table.  
The RMF theory has been successful 
to reproduce the saturation properties, masses and radii 
of nuclei, and proton-nucleus scattering data 
\cite{Ser86}.
The effective interaction used in the RMF theory is checked 
by the recent experimental data of unstable nuclei 
in neutron-rich environment close to 
astrophysical situations \cite{Sug94,Hir97}.  

We stress that the RMF theory \cite{Sug94} is based 
on the relativistic Br\"uckner-Hartree-Fock (RBHF) theory 
\cite{Bro90}.
The RBHF theory, which is a 
microscopic and relativistic many body theory, 
has been shown to be successful to reproduce the 
saturation of nuclear matter starting from the 
nucleon-nucleon interactions determined by 
scattering experiments.  
This is in good contrast with non-relativistic 
many body frameworks which can account for the saturation 
only with the introduction of extra three-body 
interactions.  

The RMF framework with the parameter set TM1, 
which was determined as the best one to reproduce 
the properties of finite nuclei including n-rich ones 
\cite{Sug94},
provides the uniform nuclear matter with 
the incompressibility of 281 MeV 
and the symmetry energy of 36.9 MeV.  
The maximum neutron star mass calculated for the cold 
neutron star matter in the RMF with TM1 is 2.2 M$_{\odot}$ \cite{Sum95b}.  
The table of EOS covers the wide range of density, 
electron fraction and temperature, 
which is necessary for supernova simulations.
The relativistic EOS table has been applied to 
numerical simulations of 
%proto-neutron star cooling \cite{},
r-process in neutrino-driven winds \cite{Sum00} and 
prompt supernova explosions \cite{Sum01}, and 
other simulations \cite{Sum95c,Ros03,Sum04}.

For comparison, we adopt also the EOS by Lattimer and Swesty \cite{Lat91}.  
The EOS is based on the compressible liquid drop model for 
nuclei together with dripped nucleons.  
The bulk energy of nuclear matter is expressed in terms 
of density, proton fraction and temperature 
with nuclear parameters.
The values of nuclear parameters are chosen to be the ones 
suggested from nuclear mass formulae and other theoretical 
studies with the Skyrme interaction.  
Among various parameters, the symmetry energy is set to be 29.3 MeV, 
which is smaller than the value in the relativistic EOS.  
As for the incompressibility, 
we use the EOS with 180 MeV, 
which has been often used for recent supernova simulations.  

\section{Simulations of core-collapse supernovae}

We have developed a new numerical code of neutrino-transfer 
hydrodynamics \cite{Yam97,Yam99,Sum04a} for supernova simulations.
The code solves hydrodynamics and neutrino-transfer at once 
in general relativity under the spherical symmetry.  
The Boltzmann equation for neutrinos is solved 
by a finite difference scheme (S$_N$ method) together 
with lagrangian hydrodynamics in the implicit manner.  
The implicit method enables us to have a longer 
time step than the explicit method, therefore, our code 
is advantageous to follow long term behaviors after the core 
bounce.
We adopt 127 spatial zones and discretize 
the neutrino distribution function 
with 6 angle zones and 14 energy zones
for $\nu_e$, $\bar{\nu}_e$, $\nu_{\mu/\tau}$ and $\bar{\nu}_{\mu/\tau}$, respectively.
The weak interaction rates regarding neutrinos are based on the 
{\it standard} rates by Bruenn \cite{Bru86}.  
In addition to the Bruenn's standard neutrino processes, 
the plasmon process and the nucleon-nucleon bremsstralung 
processes are included \cite{Sum04a}.  
The collision term of the Boltzmann equation is explicitly 
calculated as functions of neutrino angles and energies.
As an initial model, we adopt the profile of iron core of 
a 15M$_{\odot}$ progenitor from Woosley and Weaver \cite{Woo95}.
We perform the numerical simulations with Shen's EOS (denoted by SH) 
and Lattimer-Swesty EOS (LS) for comparisons.  

\begin{figure}[htb]
\begin{minipage}[t]{80mm}
\epsfig{file=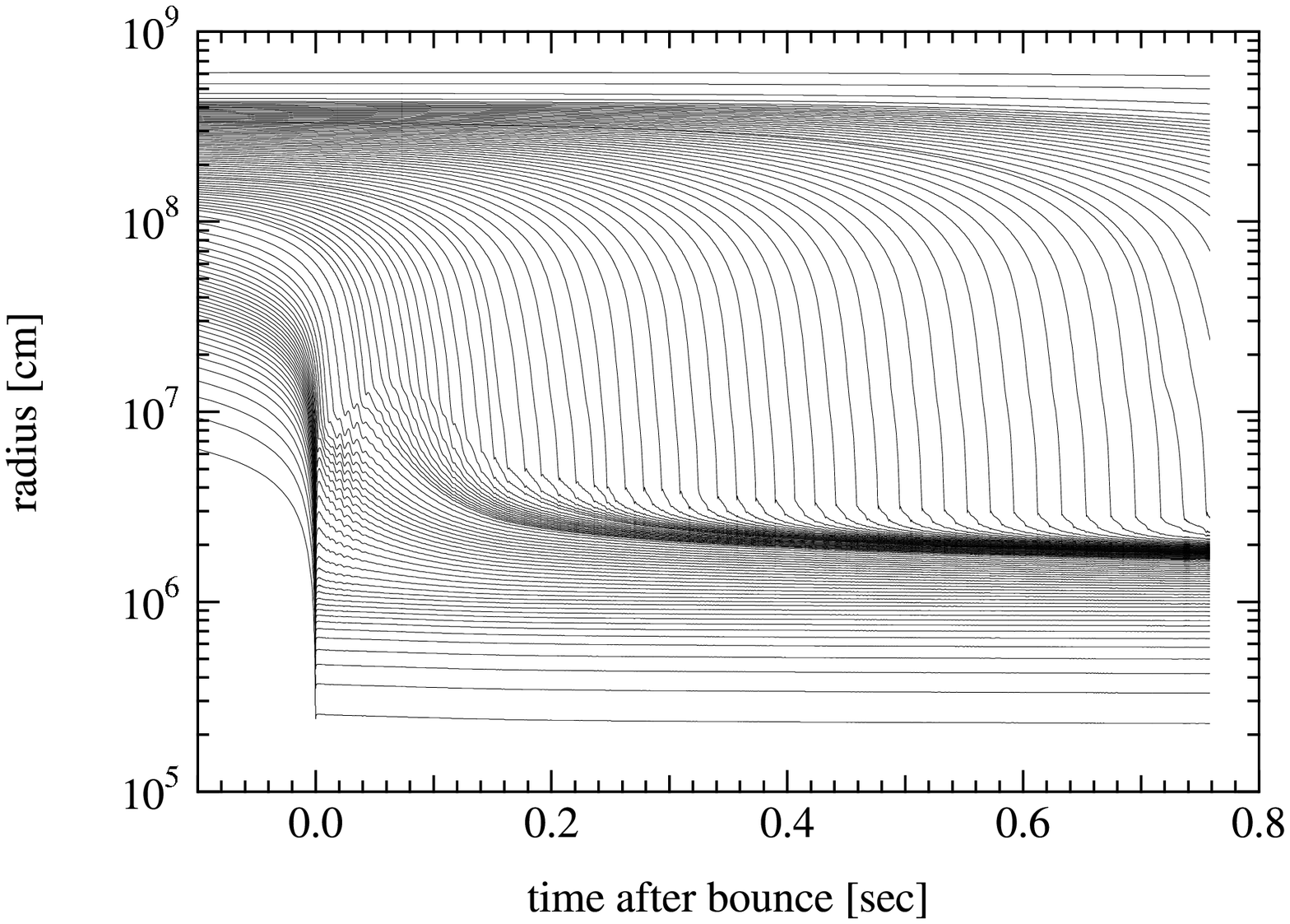,width=79mm}
%\framebox[79mm]{\rule[-26mm]{0mm}{52mm}}
\caption{Radial trajectories of mass elements of 
the core of 15M$_{\odot}$ star as a function of time after the bounce in SH.}
\label{fig:radius}
\end{minipage}
\hspace{\fill}
\begin{minipage}[t]{75mm}
\epsfig{file=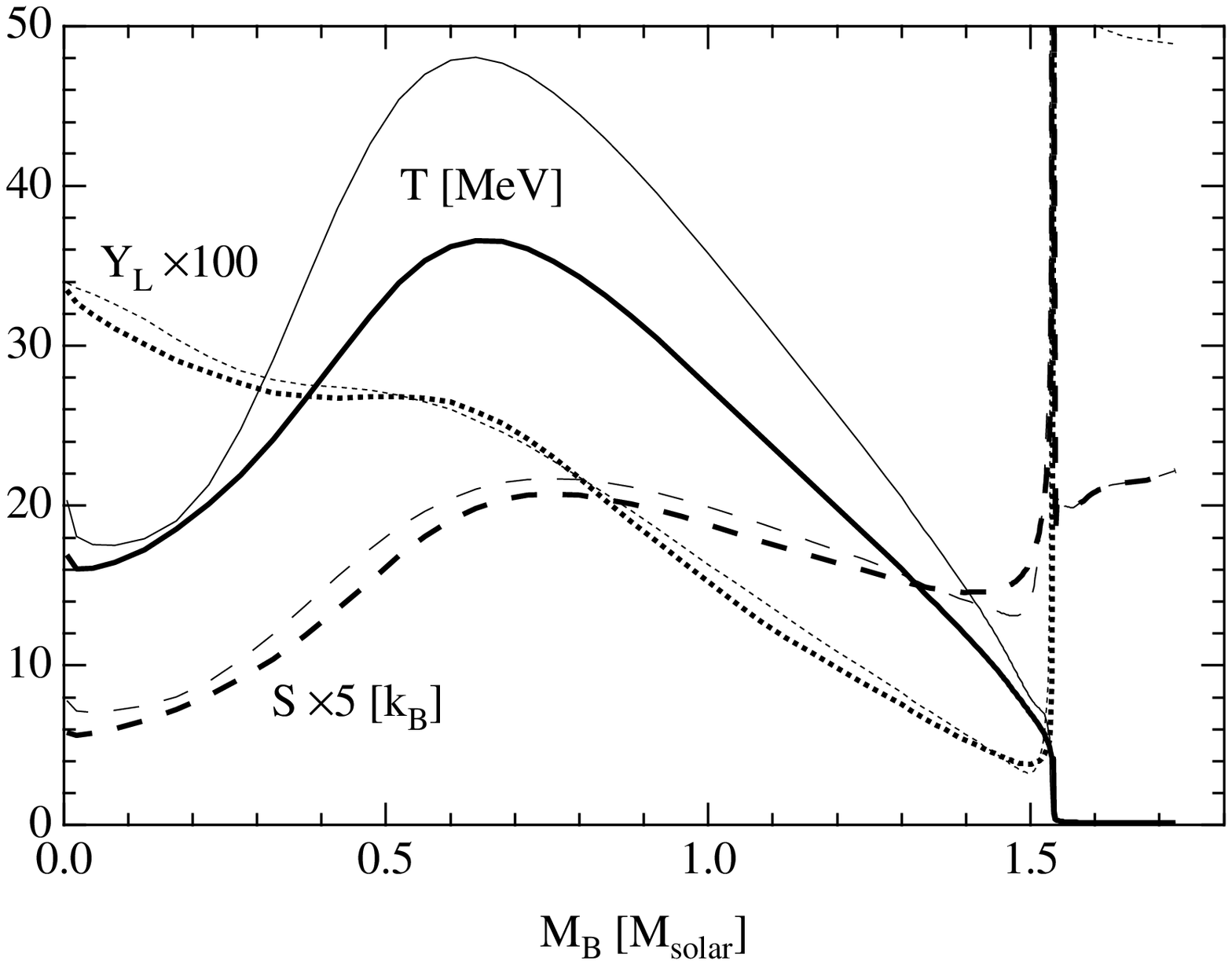,width=79mm}
%\framebox[74mm]{\rule[-26mm]{0mm}{52mm}}
\caption{Profiles of entropy, temperature and lepton fraction at 
600 ms after the bounce in SH (thick) and LS (thin).}
%$t_{pb}$=600 ms in SH (thick) and LS (thin).}
\label{fig:profile}
\end{minipage}
\end{figure}

It is remarkable that the explosion does not occur 
in the model SH, i.e. the case with the new EOS table, 
and the shock wave stalls in a similar manner as 
the model LS.  
We have found, however, that there are quantitative differences 
in core-collapse and bounce due to the differences between nuclear EOSs.  
The peak central density at the bounce in the case of SH 
is 3.3$\times$10$^{14}$ g/cm$^{3}$, which is lower than 
4.2$\times$10$^{14}$ g/cm$^{3}$ in the case of LS.  
%This is due to the difference of stiffness.
It is to be noted that, in microscopic many body calculations, 
the relativistic EOS (including Shen's EOS) 
is generally stiffer \cite{Bro90} than the 
non-relativistic EOS, on which Lattimer-Swesty EOS is based.
The stiff EOS is not advantageous in a simple argument of 
initial shock energy, however, the symmetry energy of EOS 
also plays an important role \cite{Bru89a}.
Having a larger symmetry energy, 
free-proton fractions during collapse in SH are smaller than 
in LS.  
Smaller free-proton fractions lead to smaller 
electron capture rates when electron captures on nuclei are 
suppressed.  
Because of this difference, the trapped lepton fraction at the 
bounce in SH is 0.35 at center, 
which is slightly larger than 0.34 in LS.
In the current simulations,
this difference of lepton fraction does not 
change the size of bounce core significantly. 

As a result, the shock in SH does not 
reach at a significantly larger radius than in LS (Fig. \ref{fig:radius}).  
The shock stalls below 200 km and starts receding in two cases.  
The central core becomes a proto-neutron star having 
a radius of several tens km and a steady accretion shock is formed.  
The difference of EOS becomes apparent having a more compact 
star for LS with the central density of 6.0$\times$10$^{14}$ g/cm$^{3}$ 
(3.9$\times$10$^{14}$ g/cm$^{3}$ for SH) 
 at 600 ms after the bounce.  
The peak structure of temperature at around 10 km is formed and 
the peak temperature reaches at $\sim$40 and $\sim$50 MeV %at 600 ms after bounce 
for SH and LS, respectively, 
due to the gradual compression of core having accretion (Fig. \ref{fig:profile}).  
It is noticeable that negative gradients in entropy and lepton 
fraction appear by this stage, suggesting the importance of convection.
These differences of thermal structure may give influences on 
shock dynamics, supernova neutrinos and proto-neutron star cooling.
Further details of the full numerical simulations will be published 
elsewhere \cite{Sum04a}.

\end{document}